\documentclass[]{aa} 

\usepackage{natbib}
\bibpunct{(}{)}{;}{a}{}{,} 
\usepackage{pbox}
\usepackage{url}
\usepackage{amssymb}
\usepackage[dvipsnames]{xcolor}

\usepackage{soul,xcolor}
\setstcolor{red}

\usepackage{xcolor,cancel}

\usepackage{graphicx}
\usepackage{txfonts}
\usepackage[colorlinks=true,linkcolor=blue,citecolor=blue,urlcolor=black]{hyperref}
\begin{document}

   \title{Search for gas from the disintegrating rocky exoplanet K2-22b}

   \author{A. R. Ridden-Harper \inst{1,2}
          \and I. A. G. Snellen \inst{1}
          \and C. U. Keller \inst{1} 
          \and P.  Molli{\` e}re \inst{1} 
           }
   \institute{Leiden Observatory, Leiden University, Niels Bohrweg 2, 2333 CA Leiden, The Netherlands
                  \\
              \email{arh@strw.leidenuniv.nl}            
              \and
              Department of Astronomy and Carl Sagan Institute, Cornell University, Ithaca, NY 14853, USA
              }

   \date{}

 
  \abstract
   {The red dwarf star K2-22 is transited every 9.14 hours by an object which is best explained by being a disintegrating rocky exoplanet featuring a variable comet-like dust tail. While the dust is thought to dominate the transit light curve, gas is also expected to be present, either from being directly evaporated off the planet or by being produced by the sublimation of dust particles in the tail.}
   {Both ionized calcium and sodium have large cross-sections, and although present at low abundance, exhibit the strongest atomic absorption features in comets. We therefore also identify these species as the most promising tracers of circumplanetary gas in evaporating rocky exoplanets and search for them in the tail of K2-22 b to constrain the gas-loss and sublimation processes in this enigmatic object.} 
   {We observed four transits of K2-22 b with X-shooter on the Very Large Telescope operated by ESO to obtain time series of intermediate-resolution (R $\sim$ 11400) spectra.  Our analysis focussed on the two sodium D lines (588.995 nm and 589.592 nm) and the Ca$^{+}$ triplet (849.802 nm, 854.209 nm, and 866.214 nm). The stellar calcium and sodium absorption was removed using the out-of-transit spectra. We searched for planet-related absorption in the velocity rest frame of the planet, which changes from approximately $-$66 to +66 km s$^{-1}$ during the transit.}
   {Since K2-22 b exhibits highly variable transit depths,  we analysed the individual nights and their average.  By injecting signals we reached 5$\sigma$ upper limits on the individual nights that range from 11\% - 13\% and 1.7\% - 2.0\% for the  sodium and ionized calcium absorption of the tail, respectively.  Night 1 was contaminated by its companion star so we considered weighted averages with and without Night 1 and quote conservative 5$\sigma$ limits without Night 1 of 9\% and 1.4\%, respectively.  Assuming their mass fractions to be similar to those in the Earth's crust, these limits correspond to scenarios in which 0.04\% and 35\% of the transiting dust is sublimated and observed as absorbing gas.
However, this assumes the gas to be co-moving with the planet. We show that for the high irradiation environment of K2-22 b,  sodium and ionized calcium could be quickly accelerated to 100s of km s$^{-1}$ owing to radiation pressure and entrainment by the stellar wind, making these species much more difficult to detect. No evidence for such possibly broad and blue-shifted signals are seen in our data.}
   {Future observations aimed at observing circumplanetary gas should take into account the possible broad and blue-shifted velocity field of atomic and ionized species.}


   \maketitle
   
%

\section{Introduction \label{sec:intro}}

The NASA Kepler and K2 space missions have unveiled a new class of stars which are transited in short regular intervals of a day or less by objects that are best explained as disintegrating, rocky planets.  They produce light curves that randomly vary in depth and shape (typically at $<$2\%) from one orbit to the next, showing features attributed to dust tails, such as forward scattering peaks and asymmetric transit profiles \citep[e.g.][]{Rappaport2012,Sanchis-Ojeda2015}. During some orbits, the transit can apparently be absent, implying that the transiting parent bodies themselves are too small to be detected, in line with the requirement of low surface gravities to allow a dust-tail to be launched from a planetary surface. A proposed mechanism for this mass loss is a thermally driven hydrodynamic outflow that may be punctuated by volcanic eruptions \citep{Perez-Becker2013}.  The composition of the dust likely reflects the composition of the planets, making them excellent targets to study their surface geology. For instance, mass loss and dust composition can be constrained by comparing dust-tail models to transit light curves
\citep{Rappaport2012,Rappaport2014,Brogi2012b,Budaj2013,vanLieshout2014,Sanchis-Ojeda2015,vanLieshout2016,Ridden-Harper2018,Schlawin2018} and wavelength-dependent dust extinction models to spectrophotometric observations \citep[e.g.][]{Croll2014, Felipe2013, Bochinski2015, Schlawin2016, Alonso2016}.  

However, potentially much stronger constraints on the underlying physical mechanisms of mass loss and the composition of the lost material can be derived by observing the gas that is expected to be evaporated directly from the planet or produced by the sublimation of dust particles in the tail.  The white dwarf WD 1145+017 appears to have several clumps of closely orbiting material \citep{Vanderburg2015,Rappaport2016} and was observed by \cite{Xu2016} who used Keck/HIRES to detect circumstellar absorption lines from Mg, Ca, Ti, Cr, Mn, Fe, and Ni.  \cite{Redfield2017} observed this system with Keck/HIRES and VLT/X-shooter at five epochs over the course of a year and detected varying circumstellar absorption in more than 250 lines from 14 different atomic or ionized species (O I, Na I, Mg I, Al I, Ca I, Ca II, Ti I, Ti II, Cr II, Mn II, Fe I, Fe II, Ni I, and Ni II). \cite{Izquierdo2018} made additional observations with RISE on the Liverpool Telescope and OSIRIS on the GTC.  These authors found no significant broadband wavelength dependence in transit depth and that the strong Fe II (5169\AA) circumstellar line significantly weakened during transit.  Additionally \cite{Karjalainen2019} observed the system with ISIS on the William Herschel Telescope and found a colour difference between the in and out-of-transit observations. They also found that spectral lines over the range 3800\AA\ - 4025\AA\ were shallower during transit.

\cite{Gaidos2019} searched for Na gas that was lost by the disintegrating rocky exoplanets Kepler-1520 b and K2-22 b with Subaru/HDS.  They observed one transit of Kepler-1520 b on August 1, 2014 and two transits of K2-22 b on January 26 and 29, 2016.  While they were not able to make a detection, they derived upper limits of 30\% absorption relative to the stellar spectrum.  These authors also showed with geophysical models that the amount of Na gas that is likely lost from both planets can plausibly be detected with current facilities.  Alternatively, simulations by \cite{Bodman2018}  have shown that with the James Webb Space Telescope, it should be possible to constrain the composition of dust in the tails of some disintegrating planets by directly detecting the spectral features of the dust (as opposed to the gas). 

K2-22 b is one of the four disintegrating planet systems known to date, and is the most promising for detecting gas due to its relative brightness \citep[R=15.01;][]{Rappaport2012,Rappaport2014,Sanchis-Ojeda2015,Vanderburg2015}. Its host star is an M dwarf ($T_\mathrm{eff}$ = 3830 K) that has a fainter (R = 18.79) M-dwarf companion ($T_\mathrm{eff}$ = 3290 K) approximately 2 arcsec away.  It has an orbital period of 9.146 hours and produces transit depths that vary from approximately $\lesssim$0.14 to 1.3\%, with a mean depth of 0.55\%.  The minimum transit depth implies an upper limit on the size of the disintegrating hard-body planet of 2.5 R$_\oplus$, assuming a stellar radius of 0.57 R$_\odot$ \citep{Sanchis-Ojeda2015}.

In contrast to the other known members of this class of objects, K2-22 b appears to exhibit a leading tail producing a large forward scattering peak at egress \citep{Sanchis-Ojeda2015}.   This is possible for dust particles that experience a radiation pressure force to stellar gravitational force ratio ($\beta$) of $\lesssim$ 0.02.  Such particles could either have radii $\lesssim$0.1 $\mu$m or $\gtrsim$ 1 $\mu$m.  In contrast, the post-transit forward scattering bump requires particle sizes of approximately 0.5 $\mu$m.  

A wavelength dependence in transit depth has been observed on at least one occasion \citep{Sanchis-Ojeda2015}, which allowed the Angstrom exponent, $\alpha$, to be computed, which is defined as $-d \ln \sigma / d \ln \lambda$, where $\sigma$ is the effective extinction cross-section and $\lambda$ is the wavelength.   This indicates a non-steep power-law dust size distribution with a maximum size of approximately 0.5 $\mu$m.  Considering all of these particle size constraints, \cite{Sanchis-Ojeda2015} concluded that a large fraction of particles must have sizes of approximately 1 $\mu$m.   Assuming a high-Z dust composition, they estimate a mass-loss rate of approximately 2$\times$10$^{11}$ g s$^{-1}$. 

In a large ground-based observing programme, \cite{Colon2018} observed 34 individual transit epochs of K2-22 b, of which they detected 12.  They found that the transit depths varied at a level that was consistent with the findings of previous observations.  Additionally, their data indicate some transit-like variability outside the transit window defined by the ephemeris of \cite{Sanchis-Ojeda2015}. While \cite{Colon2018} did not find strong evidence of a wavelength dependence in transit depth, their data suggest slightly deeper transits at bluer wavelengths.

In this paper we report on a search for gaseous sodium and ionized calcium in intermediate-resolution spectroscopic time-series data from VLT/X-shooter, focussing on the sodium D lines and the ionized calcium infrared triplet lines.  These species and lines were detected in WD 1145+017 by \cite{Redfield2017}, which is expected owing to their low sublimation temperatures \citep[e.g.][]{CRCHandbook}, likely presence in terrestrial planet compositions, and large absorption cross-sections \citep[e.g.][]{Mura2011}. Our study involves a lower spectral resolution than that of \cite{Gaidos2019}, and our individual exposures are also shorter (213 s versus 900 s), resulting in significantly less smearing of potential planet signals because of the change in the radial component of the orbital velocity during exposures.
This paper is structured as follows: Section \ref{sec:obs} describes our observational data, Sections \ref{sec:method} and \ref{sec:injection} describe our methods, Section \ref{sec:results} presents and discusses our results, and Section \ref{sec:conclusions} concludes. 

\section{Observational data \label{sec:obs}}

We observed transits of the rocky disintegrating planet K2-22 b on the nights of March 18 and April 4, 2017, and March 10 and March, 18, 2018 with X-shooter \citep{Vernet2011}, installed at the Cassegrain focus of ESO's Very Large Telescope Telescope (VLT) at the Paranal Observatory under programme ID 098.C-0581(A) (PI:Ridden-Harper).  The three-arm configuration of X-shooter, ultraviolet-blue (UVB), visual-red (VIS), and near-infrared (NIR) allows it to quasi-simultaneously observe the spectral range between 300 nm$ -$ 1500 nm.   

To allow the infrared background to be accurately subtracted\footnote{For a description of the background subtraction algorithm, see: \url{https://www.eso.org/observing/dfo/quality/XSHOOTER/pipeline/xsh_scired_slit_nod.html}}, these observations were carried out by nodding the telescope along the slit between two positions, A and B, in an ABBA pattern, where A and B were separated with a nod throw length of 4 arcsec.  During the three hours of observations on each night, 26 individual exposures of 213 seconds were obtained in the VIS arm. The observing dates, transit timing, exposure times, and orbital phase coverage are shown in Table \ref{table:timing}.  We used slit widths in the UVB, VIS, and NIR arms of 0.5, 0.7, and 0.4 arcsec, which resulted in resolving powers of R $\approx$ 9700, 11400, and 11600, respectively.  The physical pixels sizes in each arm, in the same respective order, are 15 $\mu$m, 15 $\mu$m, and 18 $\mu$m, which correspond to 2.9, 4.5, and 8.4 pixels per resolution element.  

X-shooter does not have an atmospheric dispersion corrector (ADC). Therefore after every hour of observing the target was re-acquired and the slit was aligned again to the parallactic angle to minimize slit losses.  The observations were reduced using the standard nodding mode recipes from the X-shooter Common Pipeline Library (CPL) \footnote{Available at: https://www.eso.org/sci/software/pipelines/xshooter/}. To enable sky background subtraction, every two exposures (AB or BA) were combined, resulting in 13 1D wavelength-calibrated spectra. 

The spectra of Night 1 were affected by time variable contamination from the faint M-dwarf companion of K2-22 that moved out of the slit.  Because of the apparent difference in spectral type between the target and the companion, the observed depth of the stellar absorption lines changes, making accurate relative spectrophotometry challenging. We therefore carried out analyses that included and excluded Night 1.

\begin{table*}[h]
\tiny
\centering
\caption{Details of the observations.  The orbital phases of K2-22 b are based on the orbital parameters derived by \cite{Sanchis-Ojeda2015}.}
\label{table:timing}

\begin{tabular}{l l l l l}
\hline \hline 
\pbox{30cm}{data set} & \pbox{20cm}{Night 1} & \pbox{20cm}{Night 2} & \pbox{20cm}{Night 3} & \pbox{20cm}{Night 4} \\

\hline 

\pbox{30cm}{date (UTC)} & \pbox{20cm}{19 Mar `17} & \pbox{20cm}{4 Apr `17} & \pbox{20cm}{11 Mar `18} & \pbox{20cm}{19 Mar `18}\\

\pbox{30cm}{start phase}  & 0.832 & 0.833 & 0.856 &  0.806\\  
\pbox{30cm}{end phase}  & 0.150 & 0.147 & 0.178  & 0.122 \\  
\pbox{30cm}{cadence (s)}  & 419.3 & 414.3  & 423.2 & 416.6  \\ 
\pbox{30cm}{exposure time (s)} & 213$^\dagger$ & 213$^\dagger$ & 213$^\dagger$ & 213$^\dagger$\\ \hline

\pbox{30cm}{observation start (UTC)} & 02:00:34 & 02:08:22 & 03:54:45 & 03:30:56 \\  
\pbox{30cm}{transit start* (UTC)} & 03:00:31 & 03:08:45 & 04:41:20 & 04:45:12\\
\pbox{30cm}{mid-transit time* (UTC)} & 03:24:31 & 03:32:45 & 05:05:20 & 05:09:12  \\ 
\pbox{30cm}{transit end* (UTC)} &  03:48:31 & 03:56:45 & 05:29:20 & 05:33:12  \\
\pbox{30cm}{observation end (UTC)} & 04:55:18 & 05:01:00 & 06:51:06 & 06:24:31 \\ \hline

\pbox{30cm}{Nr. spectra pre-transit} & 5 & 5 & 4 & 6 \\ 
\pbox{30cm}{Nr. spectra in transit} & 4 & 4 & 4 & 4  \\ 
\pbox{30cm}{Nr. spectra post-transit} & 4 & 4 & 5 & 3 \\ 
\pbox{30cm}{total Nr of spectra} & 13 & 13 & 13 & 13 \\ \hline 
Na D line region  S/n.$^1$ &  &  &  &  \\
\pbox{30cm}{average S/n. per spectrum} &  30.79 & 38.44 & 35.65 & 27.92 \\
\pbox{30cm}{total S/n} & 61.58 & 76.89 & 71.30 & 55.84 \\
Ca$^+$ triplet region  S/n.$^2$ &  &  &  &  \\
\pbox{30cm}{average S/n. per spectrum} & 101.17 & 111.27 & 113.65 & 111.34 \\
\pbox{30cm}{total S/n} & 202.33 & 222.54 & 227.29 & 222.69 \\
\hline
\pbox{30cm}{Na 5$\sigma$ limit} & 12\% & 11\% & 14\% & 13\%  \\
\pbox{30cm}{Ca$^+$ 5$\sigma$ limit} & 1.8\% & 1.7\% & 1.7\% & 2.0\% \\
\hline
\multicolumn{5}{l}{* The transit times are barycentric adjusted to be as measured at the observatory.} \\
\multicolumn{5}{l}{$\dagger$ Except for the last two spectra which have exposure times of 211 s.} \\
\multicolumn{5}{l}{1 and 2: derived from the residuals after dividing by the mean spectrum of the featureless} \\
\multicolumn{5}{l}{regions 5961.0 \AA\ $-$ 5965.2 \AA\ and 8584.8 \AA\ $-$ 8591.8 \AA, respectively.}
\end{tabular}

\end{table*}

\begin{figure*} 
\centering 
\includegraphics[width=0.8\textwidth]{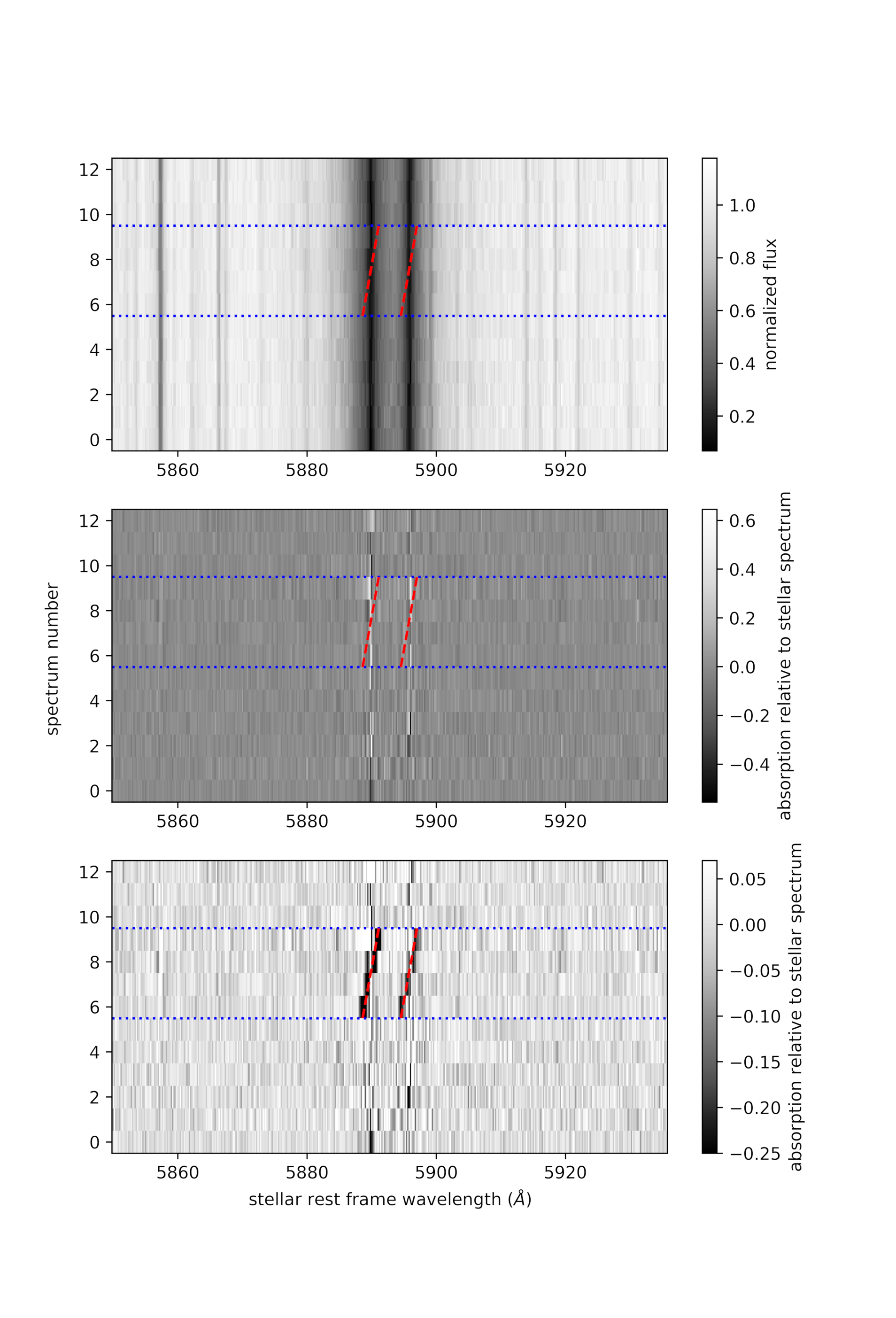}
\caption{Visual representation of the processing steps as described in Section \ref{sec:method}.  This figure shows the data of Night 4, but the other nights are very similar.  The vertical axis of each matrix represents the sequence number of the observed spectrum. The first panel shows the data around the sodium D lines after normalization and alignment in Step 2. The second panel shows the residual matrix after dividing through the average star spectrum and subtracting the mean (Step 4). The bottom panel shows the data after injecting an artificial planet signal before Step 2 that absorbs 50\% of the stellar flux. The injected planet signal can be seen as a diagonal trace (following the red dashed lines) from spectrum number 6 to 9 (indicated by the blue dotted lines), resulting from the change in the radial component of the planet orbital velocity during transit.}
\label{fig:ImageMatricesNaD} 
\end{figure*}

\begin{figure*} 
\centering 
\includegraphics[width=0.8\textwidth]{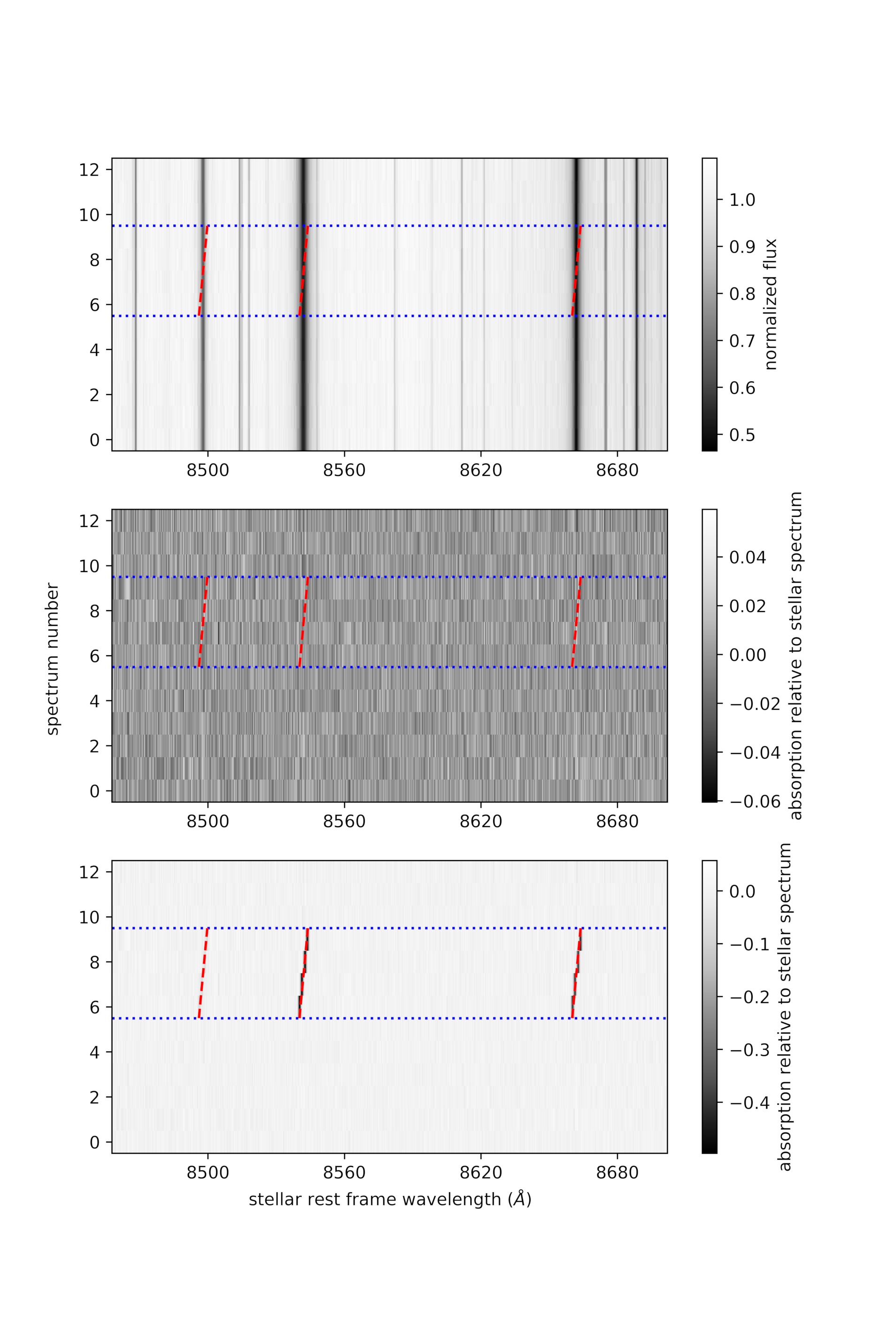}
\caption{Same as Fig. \ref{fig:ImageMatricesNaD} but for the ionized calcium NIR triplet.}
\label{fig:CaTripImageMatrices} 
\end{figure*}

\section{Analyses \label{sec:method}}

Our analyses focussed on the two sodium D lines (588.995 nm and 589.592 nm) and the Ca$^{+}$ NIR triplet (849.802 nm, 854.209 nm, and 866.214 nm), which were both captured by the VIS arm of X-shooter. The observed spectral regions are dominated mainly by stellar and some telluric lines. Both the reflex motion of the star around the barycentre of the system\footnote{\cite{Sanchis-Ojeda2015} did not detect any radial velocity variations in the spectrum of K2-22 (accurate to $\pm$ 0.3 km s$^{-1}$).} and the change in the radial component of the velocity of the observatory towards the star are so small that they can be considered to be non-variable; the position\footnote{\cite{Figueira2010} found that telluric lines are stable to 10 ms$^{-1}$ (rms).} (but not necessarily the strength) of the telluric lines is also considered non-variable. In contrast, the orbital velocity of the planet is large, leading to a change in the radial component during transit from approximately $-$66 to +66 km s$^{-1}$.  The resulting Doppler shift of the planetary lines can be used to separate them from the stellar and telluric features.  The analysis was carried out as in \cite{Ridden-Harper2016}, but is summarized below for completeness. It is comprised of the following steps and is nearly identical for the investigation of both the calcium and sodium lines.  

\begin{enumerate}

\item Normalization to a common flux level:  Variable slit losses and atmospheric scattering cause the spectra to have different flux levels. This is corrected by division through their median value over a wavelength range from 5810.8 \AA\ to 5974.6 \AA, which is close to the targeted lines (to avoid offsets due to variable low-frequency trends in the spectra). Normalizing based on shorter intervals that are centred on the targeted lines or located entirely at shorter or longer wavelengths does not change the results.  This normalization is possible because transmission spectroscopy depends on the relative change in flux as a function of wavelength and is therefore not an absolute measurement.

\item Alignment of the spectra: Owing to instrumental instability, the wavelength solution is prone to changes at a subpixel level. Since the absolute wavelength solution is not relevant for our analysis we did not explicitly measure or correct for the systemic velocity of the system. Instead, the positions of strong spectral lines are fitted in each spectrum and used to shift all spectra to a common wavelength frame.    

\item Removal of cosmic rays: Cosmic rays were removed by searching for 5$\sigma$ outliers and replacing them with a value interpolated from a linear fit to the other spectra at the affected wavelength position. 

\item Removal of stellar and telluric lines: All stationary spectral components in the spectra were removed by dividing every pixel in a spectrum by the mean value of the out-of-transit spectra at that wavelength position during the night.  Since the Doppler shift of the planet lines changes by approximately 5 pixels during the transit, this procedure has only a limited effect on potential planet lines.  Variability in the strengths of telluric lines can complicate the removal of telluric lines (see below).  However, as shown in Figs. \ref{fig:NaD1D} and \ref{fig:CaTrip1D}, the telluric lines did not exhibit significant variation. 

\begin{figure*}
\centering 
\includegraphics[width=0.7\textwidth]{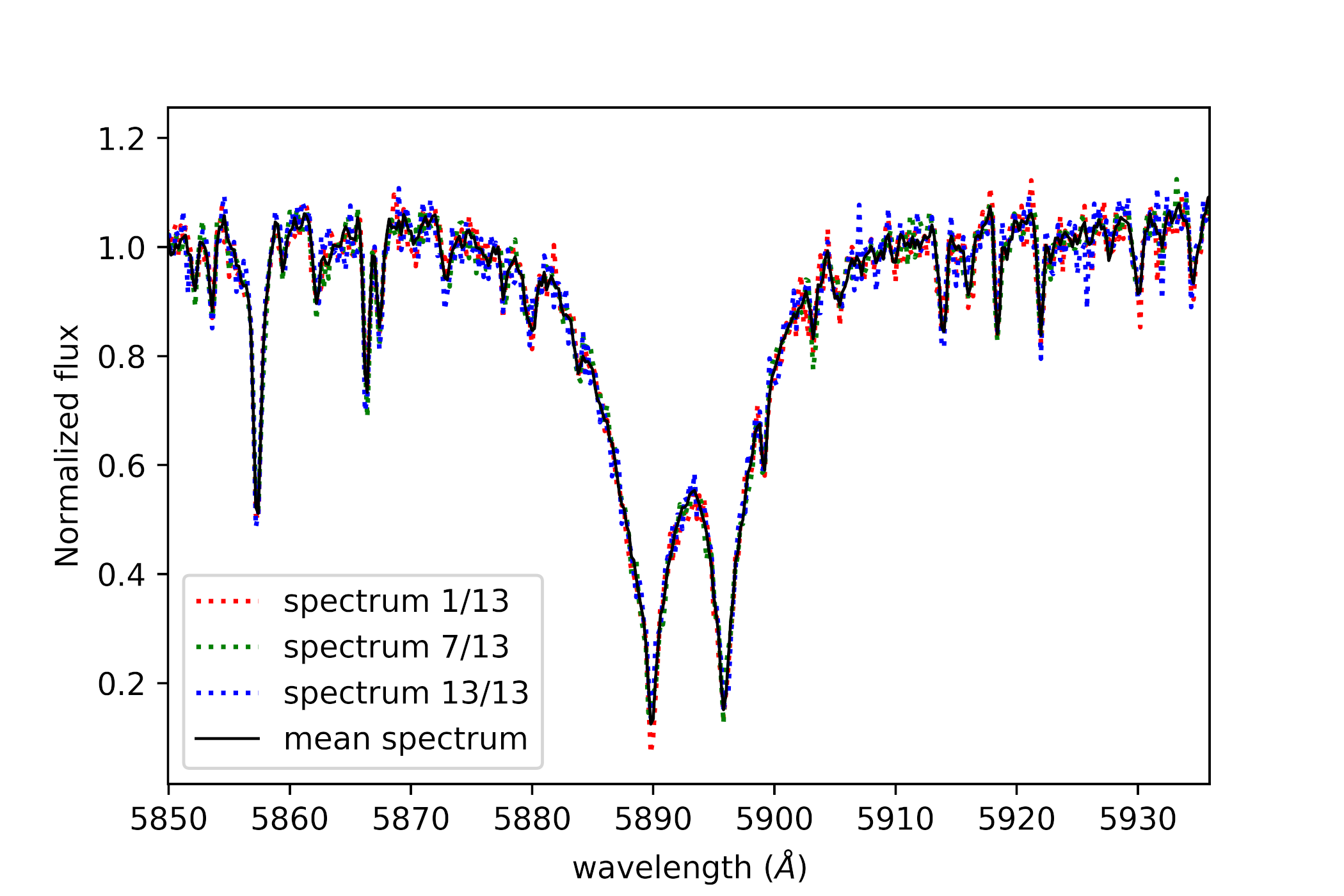} 
\caption{Region around the Na D lines used in this analysis from Night 4.  The red, blue, and green dotted lines represent the first, middle, and last observed spectra, and the black solid line represents the mean spectrum.} 
\label{fig:NaD1D} 
\end{figure*}

\begin{figure*}
\centering 
\includegraphics[width=0.8\textwidth]{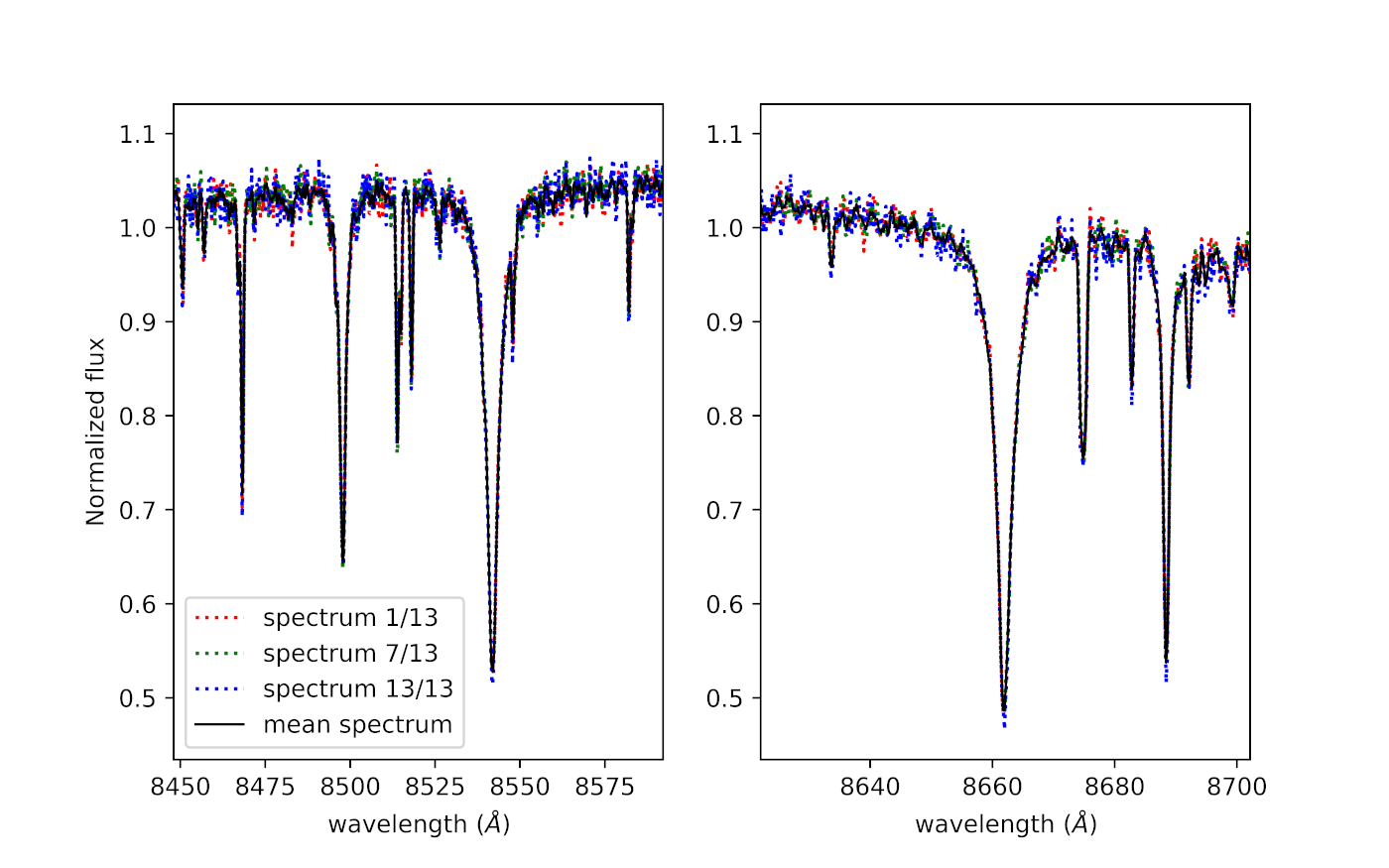} 
\caption{Same as Fig. \ref{fig:NaD1D} except for the region around the Ca$^+$ infrared triplet lines used in this analysis.} 
\label{fig:CaTrip1D} 
\end{figure*}

\item Down-weighting of noisy parts of the spectrum: Noisy parts of the spectrum, for example in the centre of strong absorption lines or telluric lines can have a significant effect on the cross-correlation functions. Therefore the flux points at each wavelength are weighted down according to the signal-to-noise ratio (S/N)\ as a function of wavelength, derived from the standard deviation of the residual spectra at each wavelength position.  This function was scaled differently for each spectrum, such that the wavelengths at which the signal of the planet is expected to be located were not changed (i.e. scaled by a factor of one).  This preserved the fractional absorption of the injected planet signal relative to the stellar spectrum and scaled the noise at wavelengths far from the cores of the strong stellar lines such that it became comparable to the noise in the cores.  This effectively weighed down spectra where the spectral lines of the planet overlap with the cores of the stellar lines when they are used to make the final 1D spectrum of that night. The expected location of the signal of the planet was calculated using the transit ephemeris from \cite{Sanchis-Ojeda2015} and assuming the signal shape to be Gaussian with a width comparable to the change in the radial velocity of the planet during the exposure (as with the injected signals in Section \ref{sec:injection}).    

\item Combination of individual lines.  The data from the three individual ionized calcium lines were combined after weighting by the line strengths.  The two sodium lines were combined in the same way.  

\item Combining the individual nights.  We shifted the line-combined spectra to the planet rest frame using the transit timing parameters from \cite{Sanchis-Ojeda2015}, and subsequently summed over all spectra taken during transit. These 1D spectra were subsequently combined for the different nights, weighted by their average S/N during the night (See Table \ref{table:timing}).
\end{enumerate}

In many other data sets, variable telluric lines cause structures in the residual spectra that can be removed with a principle component analysis (PCA), which involves decomposing the data into principle components and removing the dominant structures by subtracting the first few dominant components \citep[e.g.][]{Ridden-Harper2016}.  However, the telluric lines did not significantly vary during these observations (as shown in Figs. \ref{fig:NaD1D} and \ref{fig:CaTrip1D}) so PCA did not improve the recovery of our injected signals (see Section \ref{sec:injection}) and was not applied.  

Additionally, the Rossiter-McLaughlin (RM) effect \citep{Rossiter1924, McLaughlin1924, Queloz2000} often needs to be corrected for \citep[e.g][]{Brogi2016}, as it can distort the stellar lines during transit and induce a variable spectroscopic signal that can contaminate the transmission spectrum of the planet.  However, we do not expect the RM effect to be significant in this case, as it is more important at high spectral resolutions and for fast-rotating stars, while our data is of medium resolution and K2-22 is a slow rotator, with rotation period of 15.3 days \citep{Sanchis-Ojeda2015}.
Visual representations of the data analysis process for both the Na D doublet and the Ca$^+$ triplet are shown in Figs. \ref{fig:ImageMatricesNaD} and \ref{fig:CaTripImageMatrices}, respectively.

\section{Synthetic planet signal injection \label{sec:injection}}

Synthetic planet signals were injected after stage two of the \mbox{analysis} process (see above) to examine to what extent the analysis affects a potential planet signal and to assess the overall sensitivity of the data. The data with the artificial signals were treated in the same way as the unaltered data sets. 

We injected a simple model of the Ca$^+$ infrared triplet and the two Na D lines with relative line intensities approximated using Eq. 1 in \cite{Sharp2007}, for now ignoring terms that relate to the energy level population (e.g. temperature and partition function).  This means that the degeneracy factor, $g$, is not included in these calculations because it is part of the level population terms. This approach assumes that the population of Na atoms is in the ground state and that all of the Ca$^+$ ions are in the lower state of the triplet transition studied in this work.  For Ca$^+$ this is not the case, and we adjust the mass limits derived for this ion using its expected population statistics in Section \ref{sec:GassMassLimit}.

We took the quantum parameters that describe the line transitions from the National Institute of Standards and Technology (NIST) Atomic Spectra Database \citep{NIST_ASD}. The values and references are shown in Table \ref{tab:LineParameters}.  For the sodium D lines at 5889.95 \AA\ and 5895.92 \AA, we derive a line ratio of 2.0. For the ionized calcium triplet lines at 8498.02 \AA,\, 8542.09 \AA,\ and 8662.14 \AA, the relative line strengths derived are 0.167, 1.000, and 0.829, respectively. 

During an exposure of 213 seconds, the radial component of the orbital velocity of the planet changes by approximately 7.5 km s$^{-1}$. Since each time two exposures are combined to generate one spectrum, this effectively results in a convolution with a boxcar function with a width of 15 km s$^{-1}$, comparable to the instrumental resolution.  We therefore inject signals with a full width at half maximum (FWHM) of approximately 15 km s$^{-1}$, resulting in these signals spanning several pixels.  The planet model spectrum was Doppler shifted to the appropriate planet velocity, assuming a circular orbit \citep{Sanchis-Ojeda2015}, and injected according to

\begin{equation}
\label{eqn:SyntheticSpec}
F'(\lambda) = {[1 - C \times F_{ model}(\lambda, v_{rad})]F_{ obs}(\lambda)}
,\end{equation}

where $F_{obs}(\lambda)$ is the observed spectrum, $F_{model}(\lambda,v_{rad})$ is the Doppler-shifted model spectrum, where $C$ is a scaling parameter that determines the amplitude of strongest line, and $F'(\lambda)$ is the resulting spectrum after injecting the synthetic planet spectrum. To determine the upper limits in the strength of the ionized calcium and sodium lines, the scaling parameter $C$ was varied to reach a signal five times larger than the noise in the combined 1D planet-rest-frame spectrum.

\section{Results and discussion \label{sec:results}}

No significant signal from neither sodium nor ionized calcium was detected. Injection of synthetic planet signals indicate that 5$\sigma$ upper limits were reached in the weighted average spectrum of Nights 2 - 4 when the strength of the strongest line ($C$ in Equation \ref{eqn:SyntheticSpec}) was set to 9\% and 1.4\% for the sodium D doublet and the Ca$^+$ triplet, respectively.  We conservatively quote 5$\sigma$ limits because a systematic noise is present at the 3$\sigma$ level that was challenging to account for properly.  The limits from considering each night individually are shown in Table \ref{table:timing}.  If the dominant form of noise in the regions where the signals were injected were shot noise, the limits from the individual nights would be a factor of $\sqrt{N}$ larger than the limits from the weighted average spectrum, where $N$ is the number of nights that were averaged.  However, the limits from the individual nights are less than a factor of $\sqrt{N}$ larger, indicating that correlated noise is present in the residuals caused by the cores of the spectral lines.  In the surrounding regions, shot noise is the dominant form of noise.

Our 5$\sigma$ limit for Na of 9\% is three times lower than the limit derived by \cite{Gaidos2019} because we used shorter exposure times, which resulted in less Doppler smearing of the potential planet signal during our exposures.  We also observed four transits while they observed two, increasing our total S/N.

\begin{figure*}
\centering 
\includegraphics[width=0.8\textwidth]{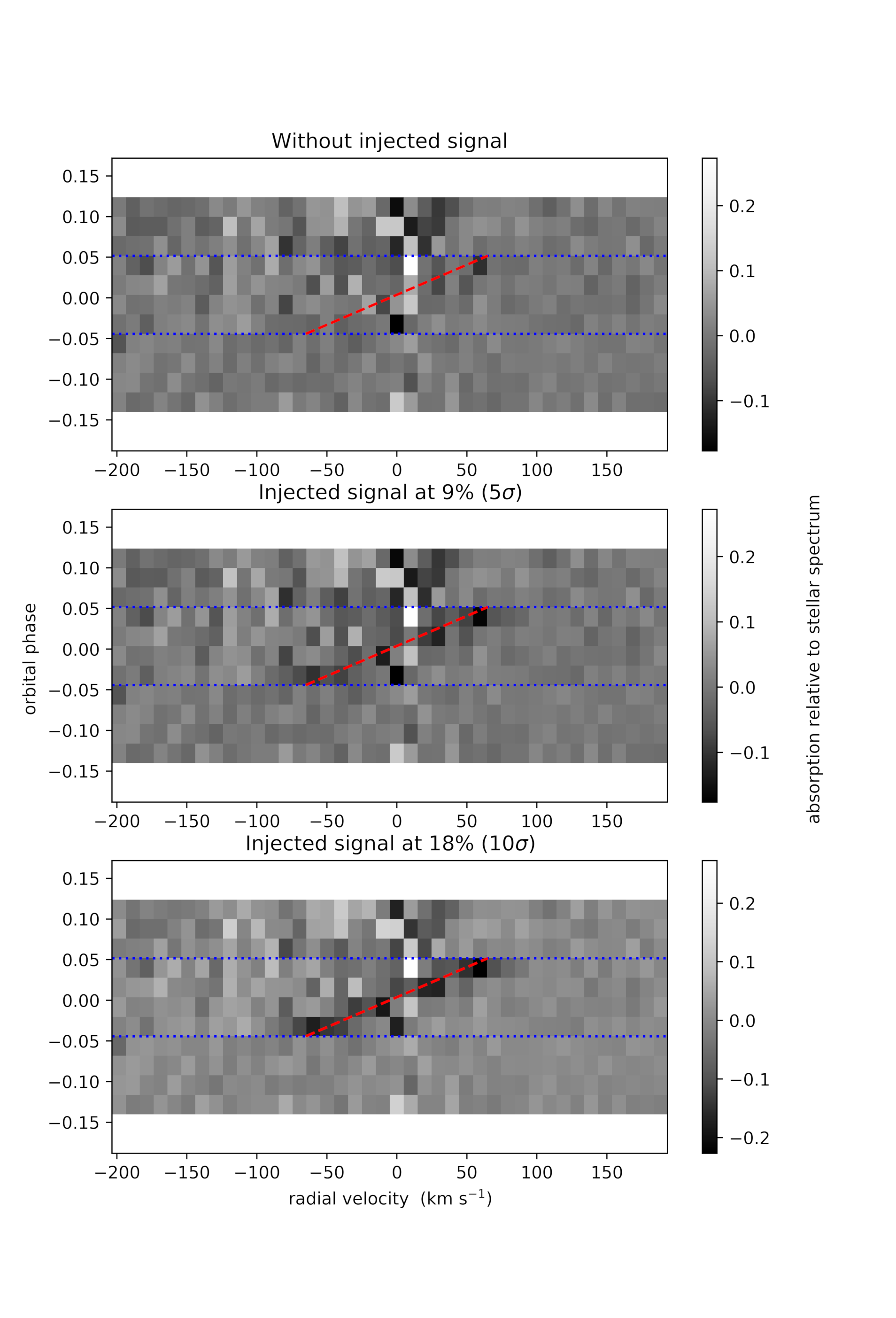} 
\caption{Planet Na D transmission spectrum combined from nights 2 $-$ 4  as a function of orbital phase (vertical axis) and radial velocity in the stellar rest frame, with no injected signal (top), a 9\% injected signal corresponding to a 5$\sigma$ limit (middle), and an 18\% injected signal (bottom).}
\label{fig:Na2Dcombined} 
\end{figure*}

\begin{figure*}
\centering 
\includegraphics[width=0.8\textwidth]{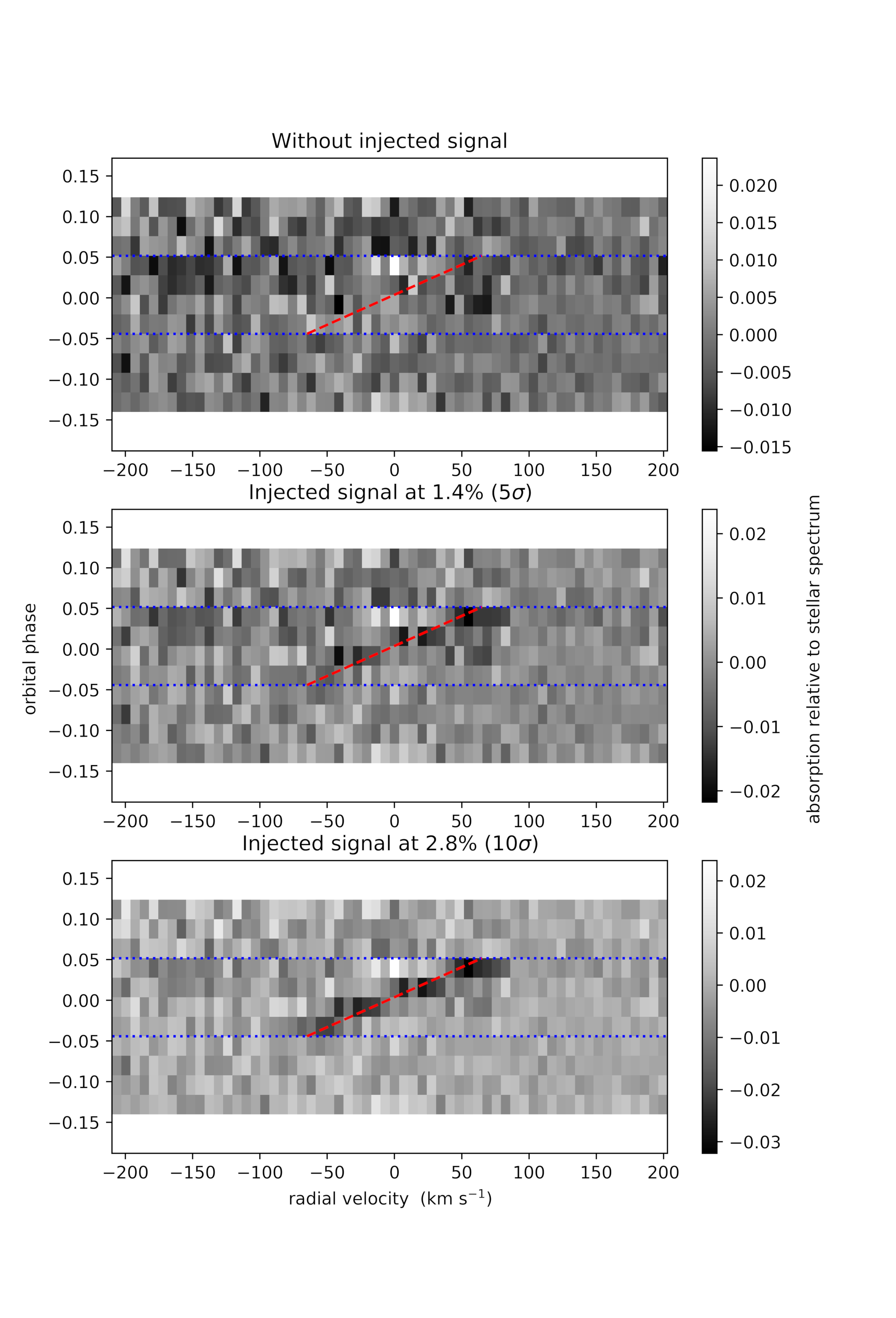} 
\caption{Same as Fig. \ref{fig:Na2Dcombined}, except for the calcium NIR triplet. The middle panel shows an injected signal of 1.4\% and the bottom panel shows an injected signal of 2.8\%.}
\label{fig:Ca2Dcombined} 
\end{figure*}

The combined (over individual lines and over nights 2 $-$ 4) transmission spectrum as a function of orbital phase is shown in Figs. \ref{fig:Na2Dcombined} and \ref{fig:Ca2Dcombined} for Na and Ca$^+$, respectively. Nearest-pixel interpolation was used, which was necessary since observations at different nights were not performed at identical orbital phases.
The panels show the data without injected signals (top), with injected signals at 5$\sigma$ (middle), and at 10$\sigma$ (bottom).  The injected signal of sodium is significantly less pronounced around mid-transit because it temporarily overlaps with the cores of the noisy stellar sodium absorption. The noise in these spectra is scaled as in Step 5 of Sec. \ref{sec:method} for the construction of the final 1D transmission spectrum.

\begin{figure*}
\centering 
\includegraphics[width=0.8\textwidth]{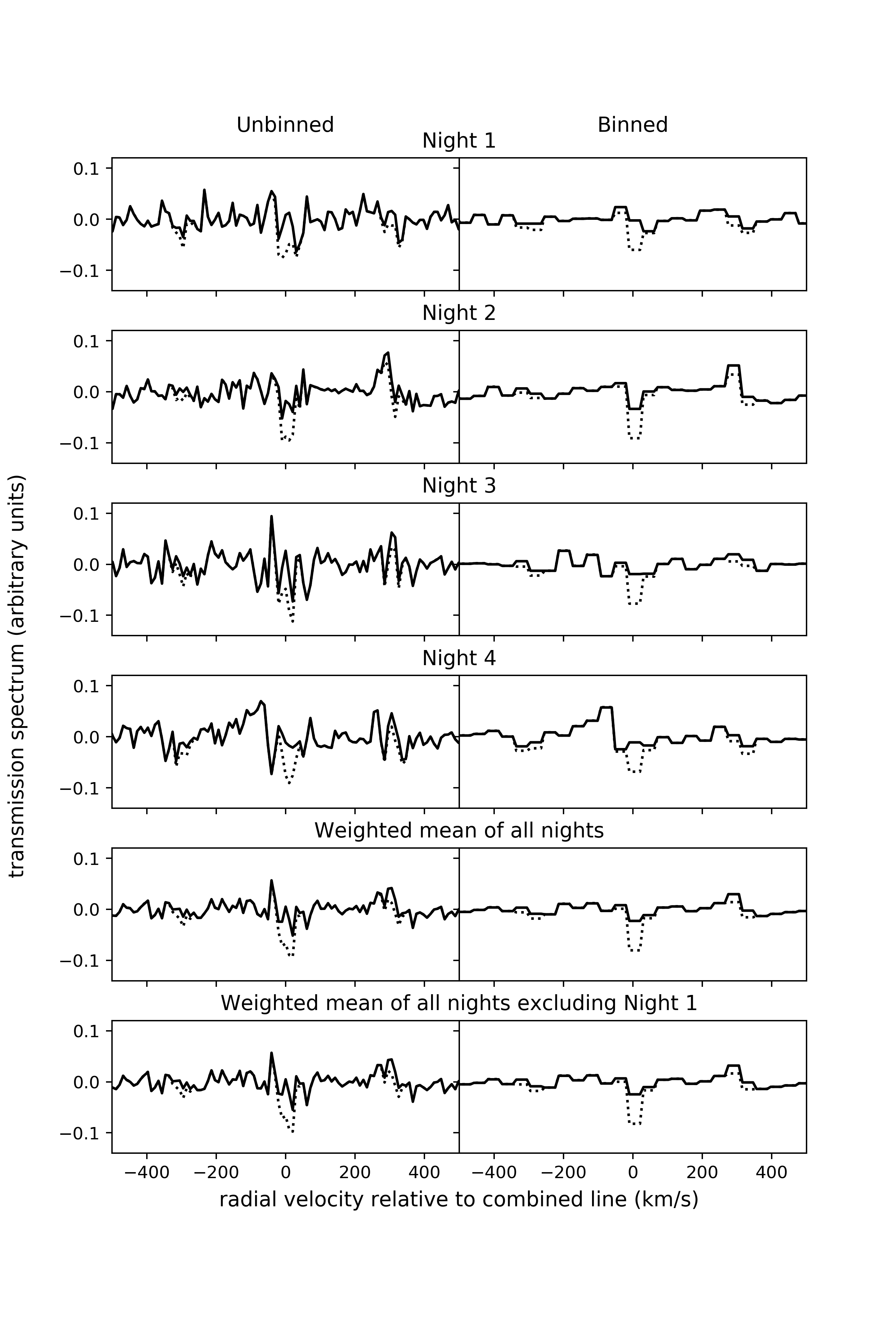} 
\caption{Planet transmission spectrum of the combined sodium D lines for all nights and their combinations with and without Night 1 (bottom panels). 
The left and right panels show the unbinned and binned (at 0.8\AA\ or 40 km s$^{-1}$) data, respectively.
The solid and dashed lines indicate the non-injected and injected data (at a 9\% depth), respectively.}
\label{fig:NaDSpec} 
\end{figure*}

\begin{figure*}
\centering 
\includegraphics[width=0.8\textwidth]{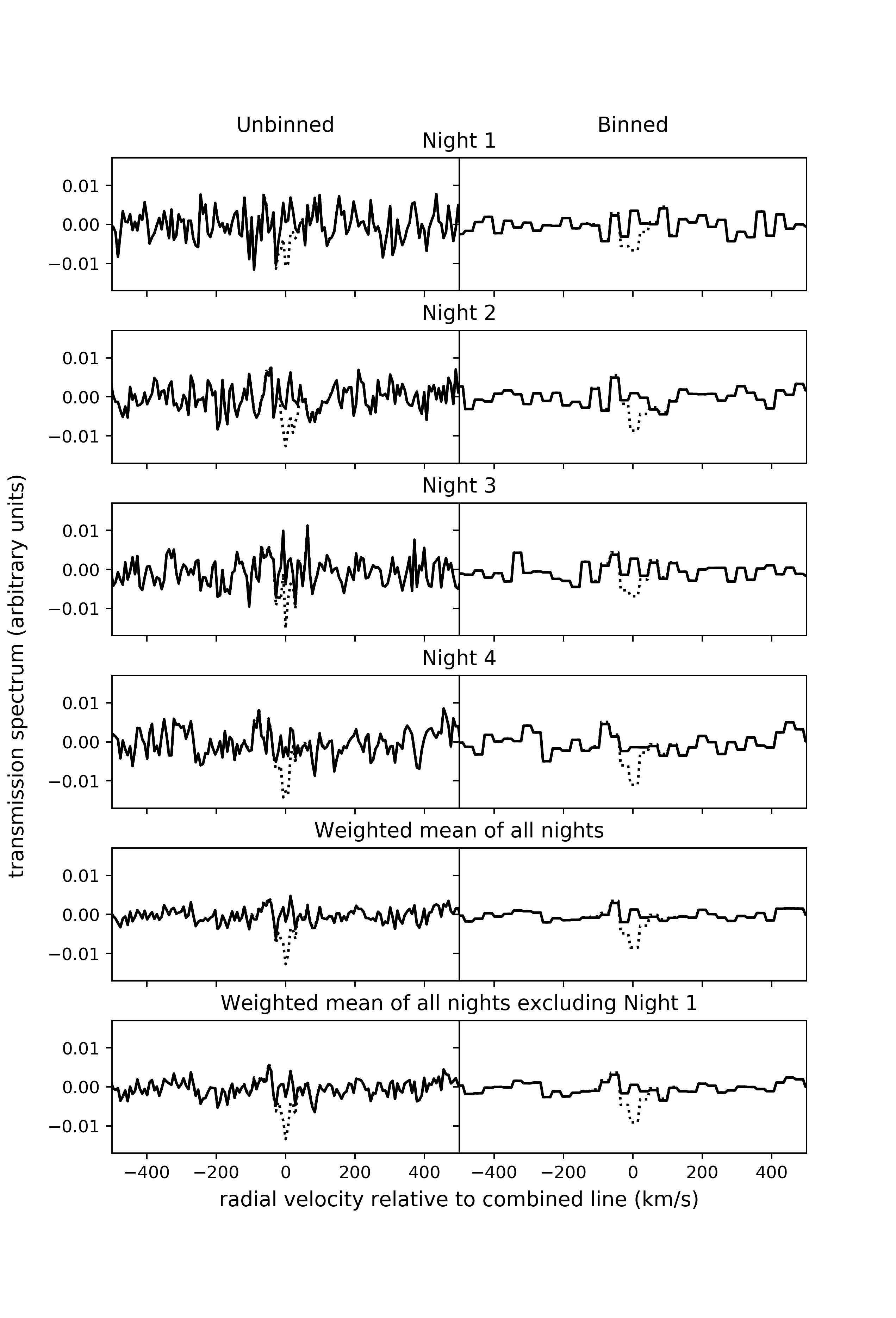} 
\caption{Same as Fig. \ref{fig:NaDSpec} but for the ionized calcium NIR triplet with an injected signal strength of 1.4\% relative to the stellar spectrum.}
\label{fig:CaTripSpec} 
\end{figure*}

The final 1D spectra per night, and those combined over all nights, and nights 2$-$4 are shown in Figs. \ref{fig:NaDSpec} and \ref{fig:CaTripSpec} (with and without the injected signals) for the Na D lines and the Ca$^+$ triplet, respectively.  The right panels are those binned by 0.8\AA\ or 40 km s$^{-1}$. Combining the two Na lines and three Ca$^+$ lines results in neighbouring lines being included in the combined 1D spectrum (e.g. for Na, the features at $\pm$ 285 km s$^{-1}$).  

\subsection{Instantaneous gas-mass limits \label{sec:GassMassLimit}}

To estimate what the observed limits mean in terms of Ca$^+$, Na, and total gas mass-loss limits, we used the following equation for absorption line strength as described in \cite{Savage1991}: 

\begin{equation}\label{eqn:opdep}
\tau(v) = \frac{1}{4 \pi \varepsilon_0} \frac{\pi e^2}{m_e c} f \lambda N(v) = 2.654\times10^{-15} f \lambda N(v)
,\end{equation}

where $\varepsilon_0$ is the permittivity of free space, $e$ is the charge of an electron, $m_e$ is the mass of an electron, $c$ is the speed of light, $f$ is the transition oscillator strength, $\lambda$ is the wavelength in Angstroms, and $N(v)$ is the integral normalized column density per unit velocity in atoms cm$^{-2}$ (km s$^{-1}$)$^{-1}$.  We assumed that lines from the gas would be too narrow to resolve at our instrument resolution of R $\sim$ 11400 so our column density profile, $N(v)$, only accounted for the natural width of the line.  This was done by setting the FWHM of a Lorentzian line profile to the natural line width  of the transition given in angular frequency units by the Einstein A coefficient (or spontaneous decay rate, $\Gamma$).  We then numerically integrated the line profile to find the required normalization factor.  The parameters that we used are shown in Table \ref{table:quantum}.   
In reality, there is also some kinetic broadening in the gas (see also the discussion below).  To approximate this without making assumptions about its pressure and temperature, we convolved the optical depth profile with the instrument resolution to broaden the line. This maximized the broadening while minimizing optical thickness effects.

\begin{table}[h]
\tiny
\centering
\caption{Spectral line transition parameters.} 

\label{table:quantum}

\begin{tabular}{c c c c c}
\hline \hline 
Spectral line \\ (in air) (\AA) & $f_{ik}$ & A$_{ki}$ (s$^{-1}$) & Normalization factor$^\dagger$ & reference \\ \hline 
\multicolumn{1}{l}{Na} & & & \\
5889.95  & 0.641 & 6.16$\times10^{7}$ & 112.03 & 1\\
5895.92 & 0.320 & 6.14$\times10^{7}$  & 112.28 & 1\\
\multicolumn{1}{l}{Ca$^+$} &  & & & \\
8498.02 & 0.0120 & 1.11$\times10^{6}$ & 4309.09 & 2\\
8542.09 & 0.072 & 9.9$\times10^{6}$ & 480.65 & 2\\
8662.14 & 0.0597 & 1.06$\times10^{7}$ & 442.69 & 2\\
\hline
\multicolumn{5}{l}{$\dagger$ Derived quantity.} \\ 
\multicolumn{5}{l}{Reference 1: \cite{Juncar1981}.} \\
\multicolumn{5}{l}{Reference 2: \cite{Edlen1956}.} \\
\multicolumn{5}{l}{Values retrieved from NIST Atomic Spectra Database \citep{NIST_ASD}.}

\label{tab:LineParameters}
\end{tabular}

\end{table}

The line intensity profile, $I(\lambda$), is estimated from the convolved optical depth profile according to  
 \begin{equation}\label{eqn:intensity}
I(\lambda) = I_0(\lambda) \exp{(-\tau(\lambda))}
,\end{equation}
where $I_0(\lambda)$ is the continuum intensity. We converted the column density $N(v)$ into a mass by assuming that the gas is optically thin and uniformly covers the stellar disk.  The 5$\sigma$ upper limit for sodium corresponds to a 9\% absorption depth for the strongest line at 5889.95 \AA, which requires an absorbing mass of sodium gas of 3.4$\times$10$^{9}$ g. This implies an upper limit on the  total gas mass of 1.4$\times$10$^{11}$ g, assuming that the composition of the dust is the same as that of the Earth's crust.  

While we assume that all sodium atoms are in the ground state and therefore can produce the targeted absorption lines, this is not the case for ionized calcium. The targeted lines originate from ions in the meta-stable $3d^2D_{3/2}$ and $3d^2D_{5/2}$ states.  To estimate the fraction of calcium II ions in these energy states, we created a simple model
consisting of three energy levels: $E_0$, $E_1$, and $E_2$. $E_0$ is the ground state ($4s^{2} S_{1/2}$), $E_1$ is the average of the $3d^2D_{3/2}$ and $3d^2D_{5/2}$ states, and $E_2$ is the average of the $4p^2 p^{0}_{3/2}$ and $4p^2 p^{0}_{1/2}$ states.  Transitions from the $E_0$ level to the $E_2$ level produce the H and K lines at 3934\AA\ \& 3963\AA, while transitions from the $E_1$ level to the $E_2$ level produce the NIR triplet lines that are probed in this study.  We also consider the classically forbidden transitions from the $E_1$ level to the $E_0$ level, which produce emission at 7291\AA\ and 7324\AA.  

Spontaneous decay from a high to low energy state occurs at a rate that is proportional to the transition's Einstein A coefficient.   Additionally, a transition from a low to high energy state occurs at a rate that is proportional to the rate of photons of energy  equal to the energy difference between the states.  We calculated the photon rates for the $E_0$ to $E_2$ and $E_1$ to $E_2$ transitions according to 

\begin{equation} \label{eqn:Nphotrate}
\gamma = \int_{-\infty}^{\infty} F_\nu \left ( \frac{R_{s}}{d} \right )^2 \frac{1}{h \nu} a(\nu) d \nu  
,\end{equation}

where $\gamma$ is the rate of photons per second, $\nu$ is the frequency, $F_\nu$ is the flux as a function of unit frequency, $R_s$ is the radius of the star K2-22, $d$ is the orbital distance of K2-22 b, $h$ is the Planck constant, and $a(\nu)$ is the cross-section of the transition, assumed to be a Lorenzitan profile of FWHM equal to the transition's natural width.  For $F_\nu$ we used a PHOENIX model spectrum \citep{Husser2013} of effective temperature $T_{\rm eff}$ = 3820 K that was normalized such that $\int_{-\infty}^{\infty} F_\nu~d\nu= \sigma T_{\rm eff}^4$, where $\sigma$ is the Stefan-Boltzmann constant.

Using these calculated transition rates, the steady-state solution of the system was found to have 0.26\% of its Ca$^+$  ions in the IR-triplet-forming $E_1$ state. For the 5$\sigma$ upper limit for Ca$^+$ of 1.4\% absorption of the  strongest line at 8542.09 \AA, the upper limits on the mass of Ca$^+$ gas and the total dust mass are 2.1$\times$10$^{12}$ g and 7.1$\times$10$^{13}$ g, respectively.

\subsection{Dust and gas mass-loss comparison \label{subsec:DustGasLossComp}}

The dust mass-loss rate required to produce the observed optical transit depth can be estimated based on the rate at which dust particles pass through the area occulting the host star.  Following the method described in \cite{Rappaport2014}, \cite{Sanchis-Ojeda2015} estimated the mass-loss rate of K2-22 b  to be 2$\times10^{11}$ g s$^{-1}$.

We can compare our derived gas-mass upper limits to the dust mass-loss rate if we assume an appropriate timescale for the absorption by the gas. We take this to be the photoionization lifetime of the absorbing species, since these species are only able to absorb at the probed transitions until they are photoionized, which on average occurs after one photoionization lifetime.  The photoionization timescale of a given species depends on both its wavelength-dependent ionization cross-section and the spectral energy distribution (SED) of the ionizing flux.  We calculated the photoionization timescales for Na and Ca$^+$ by following Eqn. \ref{eqn:Nphotrate} but only integrating over the spectral region with photon energies higher than the ionization threshold.  We used photoionization cross-sections ($a(\nu$)) taken from \cite{Verner1996} and the SED ($F_\nu$) of GJ 667C (version 2.2) from the Measurements of the Ultraviolet Spectral Characteristics of Low-mass Exoplanetary Systems (MUSCLES) survey \citep{France2016,Youngblood2016,Loyd2016}.  We used the SED of GJ 667C because its effective temperature of 3700 K is the closest match in the MUSCLES survey to K2-22's effective temperature of 3830 K.   

We find ionization lifetimes of $1.8\times10^{3}$ s and $1.1\times10^3$ s for Na and Ca$^+$, respectively.  Our Na photoionization lifetime is a factor of 3.6 shorter than the  value calculated by \cite{Gaidos2019} of 3.930$\times$10$^3$ s. They used the same MUSCLES SED, but a different Na ionization cross-section taken from \cite{Yeh1985} \& \cite{Yeh1993}.  Given the approximate nature of this estimate, the two photoionization lifetime values are sufficiently consistent. We neglected the photoionization timescale of Ca because it is only $1.8\times10^2$ s.  For comparison, the photoionization lifetimes of Na and Ca at 1 au in the solar system are $1.9\times10^{5}$ s and $1.4\times10^{4}$ s, respectively \citep{Fulle2007,Mura2011}.  For the purposes of this approximation, we do not account for the possibility of ionized Na recombining with a free electron.  

\begin{figure*}[h] 
\centering  
\includegraphics[width=0.95\textwidth]{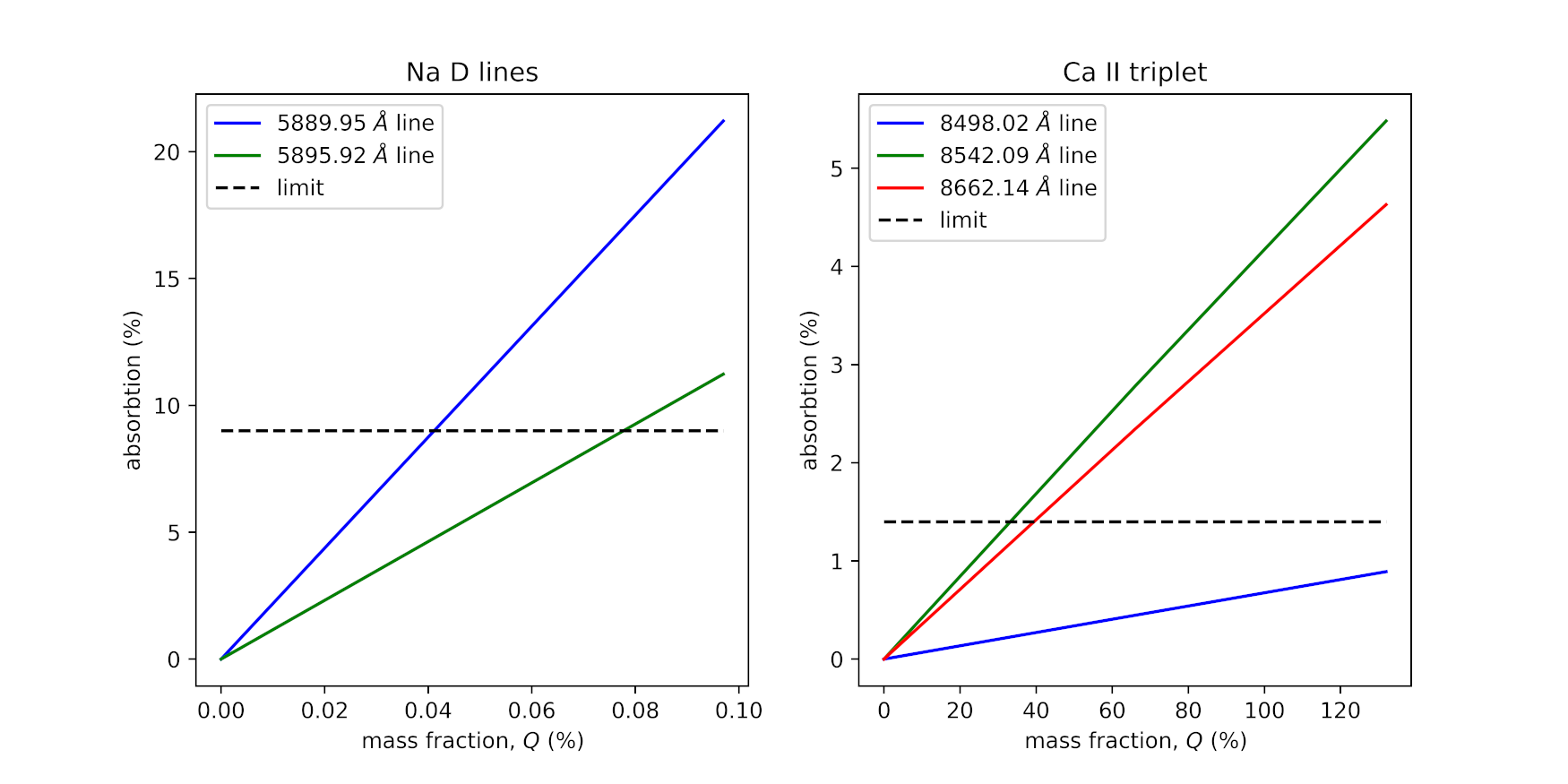}
\caption{Absorption by the Na D lines (left) and the Ca$^+$ triplet (right) as a function of fraction of available lost mass that becomes absorbing gas, assuming a dust mass-loss rate of $2\times 10^{11}$ g s$^{-1}$, Earth crust abundances and absorption lifetimes equal to the photoionization lifetimes of $1.8\times10^{3}$ s and $1.1\times10^3$ s, respectively.}
\label{fig:LineStrenghts} 
\end{figure*}

From the dust mass-loss rate and photoionization lifetimes, we can predict the expected gas column density using 
\begin{equation}\label{eqn:Mgass}
M_{gas} = Q \dot{M}_{dust} \tau_{ph}
,\end{equation}
where $\dot{M}_{dust}$ is the dust mass-loss rate of K2-22 b, $\tau_{ph}$ is the photoionization lifetime of the species, and $Q$ is the fraction of available dust mass that becomes absorbing gas for a given species.  While the sublimation rate of the dust would depend on its composition, we can assume that the dust completely sublimates after one orbit because of the seemingly random variability between consecutive transit depths.  Therefore, our $Q$ parameter cannot be used to constrain the dust composition meaningfully.  We converted the gas mass into a column density by assuming that the atoms are evenly distributed across the stellar disk.  In reality, the gas probably does not cover the entire stellar disk but this is a reasonable assumption because the gas is expected to be optically thin. The expected line strengths as a function of $Q$ are shown in Fig. \ref{fig:LineStrenghts} for the sodium D lines and the Ca$^+$ triplet.  Our conservative scaling of the photoionization timescale means that $Q$ is an upper limit.

Fig. \ref{fig:LineStrenghts} implies that if the gas were co-moving with the planet, we could have expected to detect absorption by Na and Ca$^+$ if only $Q =$ 0.04\% and 35\% of the available lost mass in dust became absorbing gas, respectively.   In contrast, it may well be that all of the dust sublimates and becomes gas ($Q = 1$) and that $Q$ may even be $>$1 because additional gas may be directly lost from the planet. It is clear that under our simplified assumptions there is no evidence for such a high $Q$ value. 

\subsection{Important Caveats: High velocity gas}

 Our estimated gas absorptions were based on important assumptions:  We assume that the dust particles completely sublimate in the time it takes them to drift across the stellar disk.  This is a reasonable assumption because the tail's exponential scale length, $l$, is estimated to be 0.19 $< l <$ 0.48 stellar radii \citep{Sanchis-Ojeda2015}.  We also assume that the gas has the same orbital velocity as the planet.  However, this may not be valid as the gas could be highly accelerated by the stellar wind and radiation pressure, giving a very broad spectral line with gas radial velocities ranging from the radial velocity of the planet to 100s of km s$^{-1}$.

\subsubsection*{Acceleration of Ca$^+$ by the stellar wind}

In the absence of a strong planetary magnetic field, ionized calcium (Ca$^+$) is dragged along by the stellar wind.  At the orbital distance of K2-22 b of 3.3 stellar radii (0.0088 au) \citep{Sanchis-Ojeda2015}, the stellar wind is still accelerating and is likely at a velocity of 60 $-$ 85 km s$^{-1}$ \citep{Johnstone2015, Vidotto2017}.  However, if the planet were to have a magnetic field, it can trap the Ca$^+$ ions, preventing these ions from being swept away by the stellar wind. This may explain the potential detection of Ca$^+$ around 55 Cancri e by \cite{Ridden-Harper2016}.    

The MESSENGER spacecraft detected Ca$^+$ in the exosphere of Mercury, however this species was trapped by the magnetic field of Mercury.  It was detected in a narrow region 2 $-$ 3 Mercury radii in the anti-solar direction, exhibiting velocities of hundreds of km s$^{-1}$.  The distribution and velocities of Ca$^+$ ions is likely due to a combination of magnetospheric convection and centrifugal acceleration \citep{Vervack2010}.

Ionized calcium has also been observed in Sun-grazing comets \citep[e.g][]{Marsden1967}. Additionally, \cite{Gulyaev2001} detected Ca$^+$ at distances of 5 $-$ 20 R$_\odot$ from the Sun and found that it had radial velocities of 170 $-$ 280 km s$^{-1}$.  These authors proposed that the Ca$^+$ is produced by the sublimation of orbiting interplanetary dust so that its final velocity is a result of its orbital motion and acceleration by the solar wind.

If the Ca$^+$ ions were swept away by the stellar wind, their spectral lines would be significantly blue-shifted and the line width would be broadened from velocities of the order of the  radial velocity of the planet, potentially up to the velocity of the stellar wind.  This would strongly hamper the detectability of this gas with the instrumental set-up discussed in this work.

\subsubsection*{Stellar radiation pressure}

Photons can exert a radiation pressure on atoms that is dependent on the wavelength-dependent photon density and the atomic absorption cross-sections.  If an atom has a radial velocity relative to the photon source, the wavelength-dependence of the absorption cross-section  causes a Doppler shift accordingly. This can have a large effect, for example for sodium, for which the stellar absorption lines can be very deep. A significant Doppler shift of these lines can increase the relevant photon flux by an order of magnitude for stellar absorption lines that are 90\% deep, causing high accelerations.

Typical accelerations of neutral sodium in the exospheric tail of Mercury are 0.2 $-$ 2 m s$^{-2}$ \citep{Potter2007}. The final velocity that an accelerated atom can reach depends on the timescale over which it is accelerated.  In our case, we are only interested in the velocity that Na and Ca$^+$ reach before they are photoionized because they stop producing the probed absorption lines when they are photoionized.  Therefore, we use their photoionization timescales as the timescale over which they are accelerated. 

\cite{Cremonese1997} observed a neutral sodium tail from comet Hale-Bopp when it was at a distance of 1 au, and measured radial velocities of sodium atoms of  60 $-$ 180 km s$^{-1}$, along its tail of sky-projected length $31\times10^6$ km. Similarly, radiation pressure accelerates hydrogen that has escaped from the evaporating atmospheres of the hot Jupiters HD 209458 b and HD 189733 b to velocities of approximately 130 km s$^{-1}$ \cite[e.g.][]{Bourrier2013b}.

The final velocity that such atoms reach in a given system is expected to be roughly independent of the distance from the host star, since the acceleration scales as $d^{-2}$, where $d$ is the orbital distance, and the ionization timescale as $d^2$, the latter counteracting the former.  

Comparing the solar spectrum with that of K2-22, using the solar absolute magnitudes from \cite{Willmer2018}; calculating the absolute magnitudes of K2-22 from \cite{Sanchis-Ojeda2015}; and considering that the mass of K2-22 is 0.6 solar masses, we find that the effective optical radiation pressure acting on the neutral sodium atoms at the location of K2-22 b is approximately 250 times higher than for the Earth in the solar system.  Additionally, our calculated photoionization timescale for sodium atoms at K2-22 b (see Sec. \ref{subsec:DustGasLossComp}) is approximately 100 times shorter than at 1 au in the solar system. 

Combining these two effects allows us to estimate, to first order, the final velocity of the accelerated Na atoms by scaling the observed solar system values.  We find a maximum velocity of approximately 250/100 = 2.5 times larger than that of sodium tails in the solar system, which evaluates to approximately 450 km s$^{-1}$. We note that potential effects of high energy activity such as flares are neglected. To estimate the velocity that Ca$^+$ ions reach after being accelerated only by radiation pressure, we use the same method but with an additional scaling to account for the different photoionization timescale and radiation pressure, which was calculated by \cite{Shestakova2015}.  We find that radiation pressure alone can cause Ca$^+$ to reach a velocity of 30\% of that of sodium in the solar system, which evaluates to 50 km s$^{-1}$.

\begin{figure}[h] 
\centering 
\includegraphics[width=9cm]{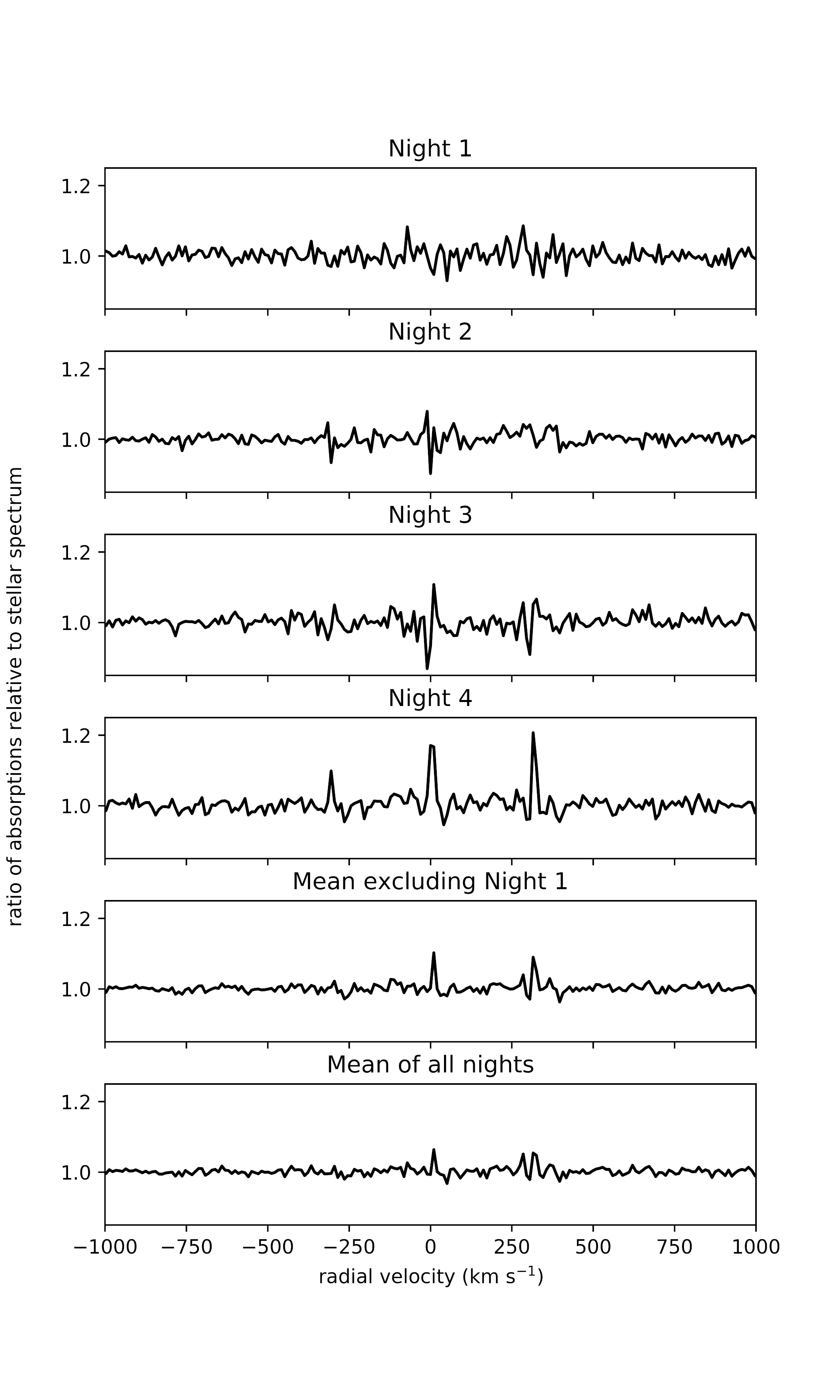}
\caption{Ratio of the average in-transit to out-of-transit signal of blue-shifted sodium gas.} 
\label{fig:NaSpecRatio} 
\end{figure}

\begin{figure}[h] 
\centering 
\includegraphics[width=9cm]{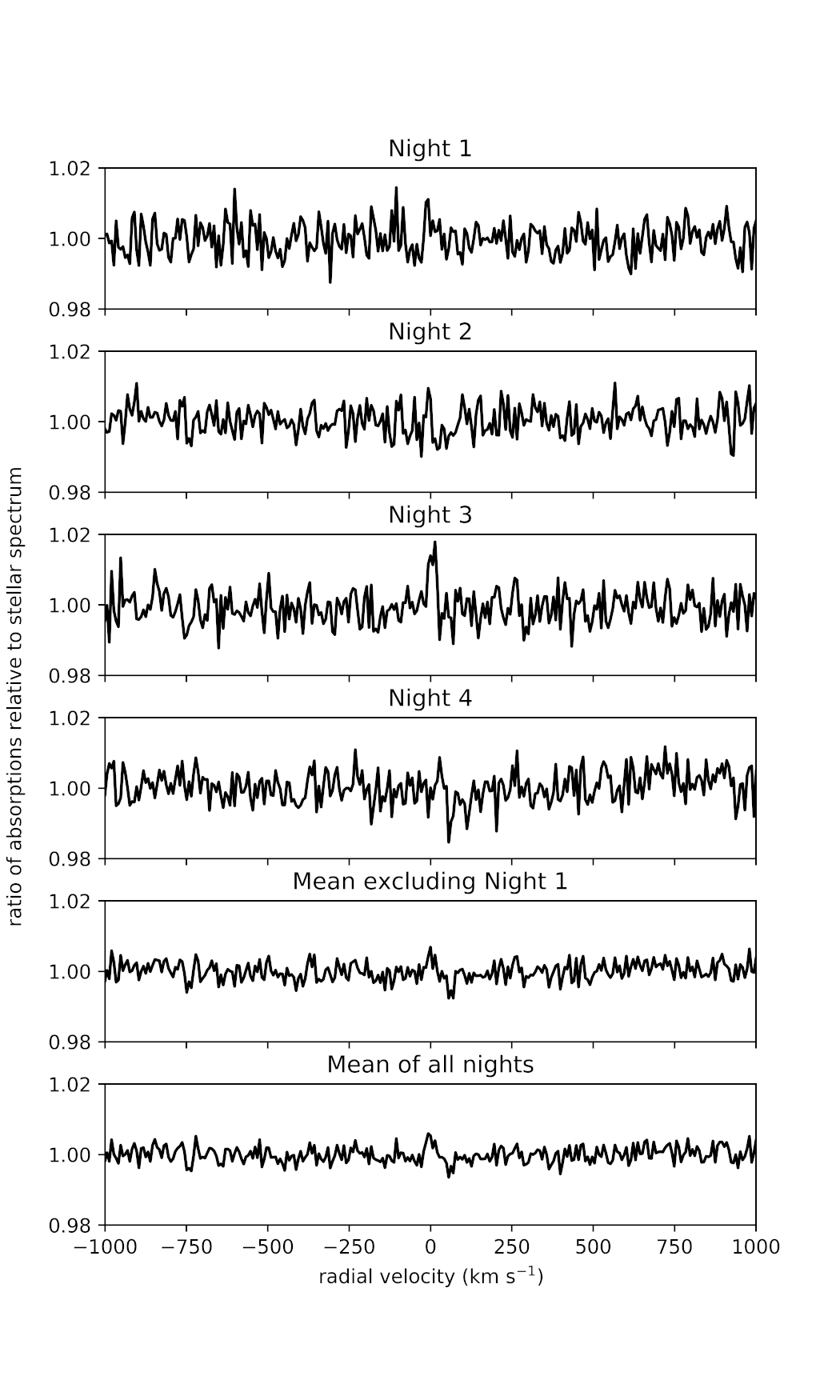}
\caption{Same as Fig. \ref{fig:NaSpecRatio} except for Ca$^+$ gas.} 
\label{fig:CaSpecRatio} 
\end{figure}

We searched for blue-shifted signals of Na and Ca$^+$ using the ratio of the average in-transit to out-of-transit signals in the residual spectra, after removing the stellar and telluric features.  Figs. \ref{fig:NaSpecRatio} and \ref{fig:CaSpecRatio} show these ratios for Na and Ca$^+$, respectively, which do not exhibit any statistically significant features over the radial velocity range of $\pm$1000 km s$^{-1}$.  There are a few outlying points but these are only due to the high noise in the cores of the targeted lines.       

While \cite{Gaidos2019} have mainly attributed their non-detection of Na around K2-22 b to their instrument resolution being too low to resolve the narrow signal they expected from the Na cloud (based on thermal broadening), and the signal from the Na cloud being blurred by the Doppler shift from the orbital motion of the planet during an exposure, they also qualitatively suggest that acceleration and shaping of the cloud by stellar winds may play a role.

\subsection{Alternative interpretations}

An alternative explanation for our non-detection is that the planet and dust particles may not have a typical terrestrial planet composition.  Furthermore, even if the planet overall has an expected composition, the dust particles may not directly reflect this.  By modelling the light curve of the similar disintegrating planet Kepler-1520 b, \cite{vanLieshout2016} found its dust composition to be consistent with corundum (Al$_2$O$_3$), which is somewhat surprising because it is not a major constituent of typical terrestrial planet compositions.  These authors suggested that this may be due to the dust grain formation process favouring the condensation of particular species or the surface of the planet being covered in a magma ocean that has been distilled to the point of containing mostly calcium and aluminium oxides.  A similar process may be occurring on K2-22 b, reducing the abundance of Na and Ca in the dust particles. 

Another potential explanation of our non-detection is that all of our observed transits happened to be during quiescent periods of low mass-loss rates.  However, based on the observed transit depth variability, we consider this to be unlikely.  It would be beneficial for future spectroscopic observations to be carried out simultaneously with optical photometric observations to allow the contemporaneous mass-loss rate to be estimated.    

\section{Conclusions and future outlook \label{sec:conclusions}}

We observed four transits of the disintegrating rocky exoplanet K2-22 b with X-shooter/VLT to search for absorption by gas that is lost directly by the planet or produced by the sublimation of dust particles in its tail.  In particular, we focussed on the sodium D line doublet (588.995 nm and 589.592 nm) and the Ca$^{+}$ NIR triplet (849.802 nm, 854.209 nm and 866.214 nm).

We detect no significant Na nor Ca$^+$ associated with the planet and derive 5$\sigma$ upper limits on their possible absorptions of 9\% and 1.4\% relative to the stellar continuum, respectively, which points to low gas-loss limits compared to the estimated average dust mass loss derived for this system. We suggest that the probed Ca$^+$ is likely accelerated to a velocity of approximately 135 km s$^{-1}$ by the combination of the stellar wind and radiation pressure (where 85 km s$^{-1}$ is due to the stellar wind), while the probed Na is likely accelerated to a velocity of 450 km s$^{-1}$ by the radiation pressure alone. This leads to very broad, blue-shifted signals, which would be hard to detect with the instrumental set-up used. We searched for such signals in our data but did not find them. 

If the signals from gas-loss are indeed very broad, it may be good to search for them using spectrographs with lower spectral resolution, either using ground-based telescopes utilizing multi-object spectroscopy for calibration, or using the future James Webb Space Telescope (JWST) $-$ although the sodium D lines are just outside the wavelength range covered by its Near-Infrared Spectrograph (NIRSpec).  In addition, other species such as  O, Mg, Ti, Cr, Mn Fe, and Ni could be searched for as they were detected in the circumstellar disk of the white dwarf WD 1145+017, which is thought to originate from disintegrating planetesimals \citep{Redfield2017}. While in principle, the combination of multiple species in the transit model would increase the chance of detection, since many lines can be combined, they may all be at a different levels of sensitivity to radiation pressure and acceleration by the stellar wind, making combination more challenging.

\begin{acknowledgements}
A. R. R.-H. is grateful to the Planetary and Exoplanetary Science (PEPSci) programme of the Netherlands Organisation for Scientific Research (NWO) for support.  I. A. G. S. acknowledges support from an NWO VICI grant (639.043.107), and from the European Research Council under the European Union’s Horizon 2020 research and innovation programme under grant agreement No. 694513.  We thank the anonymous referee for their constructive comments.  
\end{acknowledgements}

\bibliographystyle{aa} 
\bibliography{disintegrating_planets}

\end{document}